\documentclass{revtex4}
\usepackage{fullpage}
\usepackage[dvips]{graphicx}
\input{epsf}
\usepackage{amsmath}
\usepackage{amssymb}
\usepackage[mathscr]{eucal}

\newcommand{\ket}[1]{| #1 \rangle}

\def\qed{\hfill \vrule height 5pt width 5pt depth 0pt
         \medskip}

\def\proof{\noindent{\bf Proof}\ \ }
\newtheorem{theorem}{Theorem}
\newtheorem{definition}{Definition}

\newtheorem{lemma}{Lemma}
\newtheorem{corollary}{Corollary}
\newtheorem{remark}{Remark}

\begin{document}

\title{Controllability of quantum mechanical systems by root space
  decomposition of $\mathfrak{su}(N)$}
\author{Claudio Altafini}\thanks{This work was supported by a grant from the Foundation Blanceflor Boncompagni-Ludovisi.}
\affiliation{SISSA-ISAS  \\
International School for Advanced Studies \\
via Beirut 2-4, 34014 Trieste, Italy }
\email{altafini@ma.sissa.it}

\begin{abstract}
The controllability property of the unitary propagator of an $N$-level quantum mechanical system subject to a single control field is described using the structure theory of semisimple Lie algebras.
Sufficient conditions are provided for the vector fields in a generic configuration as well as in a few degenerate cases.
\end{abstract}

\maketitle 

\section{Introduction}
The question of controllability for a finite level quantum system, see Ref. \cite{Dahleh1,Schirmer1,DAlessandro2,Turinici1}, is studied in this paper by analyzing the structure of the semisimple Lie algebra of its time evolution operator.
The system dynamics is determined by its internal Hamiltonian and by an external Hamiltonian describing the interaction with a control field.
Of the several different aspects of controllability that can be defined for a closed system of such type (see Ref. \cite{Albertini1,Schirmer3} for an overview), we consider here the more direct and important in practical applications, namely the controllability of its unitary propagator which, in control terms, corresponds to a bilinear system with drift and a single control input and evolving on $SU(N)$.
For a compact semisimple Lie group like $SU(N) $, the testing of
global controllability is the simplest of all noncommutative Lie
groups. In fact, compactness implies that the accessibility property
collapses into (global) controllability and semisimplicity implies
that almost all pairs of vector fields span the corresponding
Lie algebra.
The first property means that purely algebraic tools, like the {\em
Lie algebra rank condition} normally used in control theory provides
necessary and {\em sufficient} conditions for controllability, while
the second property affirms that controllability is generically
verified even in the single control case.
The main scope of this paper is to give the interpretation of these properties in terms of structure theory of semisimple Lie algebras, see Ref. \cite{Cornwell1,Gilmore1,Sattinger1}, and to provide alternative tests to the exhaustive computation of commutators that the Lie algebra rank condition requires.
So genericity is interpreted in terms of regularity of the roots of the Lie algebra $\mathfrak{su}(N) $ and another property, regularity along the control vector field, immediately follows.
Replacing the Lie algebra rank condition means seeking for alternative conditions that guarantee the maximal nonintegrability of the pair of vector fields.
The main tool we use, together with the regularity of the roots, is the connectivity of the graph of the control vector field. Both properties were classically used to analyze controllability of vector fields on semisimple Lie algebras (especially the noncompact ones, see Ref. \cite{Jurdjevic4,Gauthier1,ElAssoudi1, SilvaLeite2}).
For the same type of problem as ours, the properties of the graph were recently used also in \cite{Turinici1}.
The conditions we obtain, based only on the {\em a priori} knowledge of the two vector fields, are only sufficient but they allows us to avoid any computation of Lie brackets.  
From the generic case, physically representing a quantum system with all different transition values between its (nondegenerate) energy levels, these tools carry on to the singular case, where some of these levels might be equispaced.
 
The paper is organized as follows: the structure theory of semisimple Lie algebras is recalled in Section \ref{sec:root-sp-dec} and it is applied to the quantum system in Section \ref{sec:qu-co-sys}, where all the needed control concepts are given.
The sufficient conditions for a generic pair of vector fields are given in Section \ref{sec:suff-co-gen}, while in Section \ref{sec:suff-co-sin} the simplest among the singular cases are analyzed.

\section{Root space decomposition for $\mathfrak{su}(N)$}
\label{sec:root-sp-dec}
Consider the classical Lie algebra $A_{N-1} $, complexification of
$\mathfrak{su}(N) $ according to Cartan's notation.
The subindex ${N-1} $ is the rank of the {\em Cartan subalgebra} $
\mathfrak{h} $ of $A_{N-1} $ 
i.e. the maximal abelian subalgebra such that the endomorphism $ {\rm
  ad}_H$ of $ A_{N-1} $ is semisimple for all $H \in \mathfrak{h} $.

An element $ H \in A_{N-1} $ is said {\em regular} if the multiplicity
of the zero eigenvalue of $ {\rm ad}_H $  is equal to the rank of $A_{N-1}$
i.e. $ {\rm dim} ( {\rm ker } \; {\rm ad } _H) = {\rm rank } A_{N-1} =
N-1 $.
The set of regular elements $H$ is open and dense in $A_{N-1} $.
Choose one such $H$ and consider the corresponding Cartan subalgebra $
\mathfrak{h} = \mathfrak{g}_0 (H) = \left\{ B \in A_{N-1} \;
  | \; {\rm ad}_H B = 0\right\} $.

The {\em roots} of $A_{N-1} $ are the functionals $ \alpha $ on $
\mathfrak{h} $ such that, for $ H \in \mathfrak{h} $, $ {\rm ad}_H B =
\alpha(H) B $, $ B \in A_{N-1} $, i.e. $\alpha $ give the
eigenvalues of $ {\rm ad}_H $ for each choice of $H$.
Denote by $ \Delta $ the set of nonzero roots of $A_{N-1} $ with
respect to $\mathfrak{h}$, by $ \Delta^+ $ the subset of positive
roots with respect to the lexicographic order on the dual of $
\mathfrak{h} $, and by $ \Phi $ the set of {\em
  fundamental roots} i.e. the set of positive roots that cannot be
written as sums of two other positive roots.

For all regular $H$, the decomposition induced by the roots has the same
structure: all nonzero eigenvalues $ \alpha (H) $ are
distinct and have multiplicity 1. $ A_{N-1} $ is decomposable into a
direct sum of {\em root 
  spaces} $ \mathfrak{g}_\alpha = \left\{ B \in A_{N-1} \;
  | \; {\rm ad}_H B = \alpha(H) B \right\} $: 
\[
A_{N-1} = \mathfrak{h} + \bigoplus _{\alpha \in \Delta} \mathfrak{g}
_\alpha 
\]
Each $  \mathfrak{g}_\alpha $ is invariant for $  {\rm ad}_H $ and
satisfies $ [  \mathfrak{g}_\alpha , \, \mathfrak{g}_\beta] =
\mathfrak{g}_{\alpha + \beta} $ where $ \mathfrak{g}_{\alpha + \beta}
= 0 $ if $ \alpha + \beta \notin \Delta$.
Furthermore, calling $K$ the Killing form, i.e. the bilinear form $ K
\; : \; A_{N-1} \times A_{N-1} \to \mathbb{R}$ , $ X, \, Y \mapsto
{\rm trace}({\rm ad }_X  {\rm ad}_Y ) $,
the restriction of $ K $ to $\mathfrak{h} $ is nondegenerate and for
each root $ \alpha \in \Delta $ there exists a unique $ H_\alpha \in
\mathfrak{h} $ such that $ \alpha(H) = K(H, \, H_\alpha )$, so that $
\alpha(H_\alpha) = K(H_\alpha , \, H_\alpha ) \neq 0 $.
If $\alpha \in \Delta $, so does $ - \alpha $ and for $ X \in
\mathfrak{g}_\alpha $ and $ Y \in
\mathfrak{g}_{-\alpha} $ $ [ X, \, Y ] = K(X, \, Y) H_\alpha $.
Therefore, by normalizing, we can choose {\em root vectors } $
E_\alpha \in \mathfrak{g}_\alpha $ such that 
\begin{equation}
\begin{split}
[ H , \, E_\alpha ] & = \alpha(H) E_\alpha \qquad \text{for $ H \in
  \mathfrak{h} $} \\
[ E_\alpha , \, E_{- \alpha } ] & = H_\alpha \\
[ E_\alpha , \, E_\beta ] & = \begin{cases}
0 & \text{ if $ \alpha + \beta \notin \Delta $} \\
N_{\alpha \beta } E_{\alpha + \beta } & \text{  if $ \alpha + \beta
  \in \Delta $}
\end{cases}
\end{split}
\label{eq:weyl-bas-1}
\end{equation}
where $N_{\alpha \beta } $ are real constants and $ N_{\alpha \beta } = -N_{( -\alpha)  (- \beta) } $
From \eqref{eq:weyl-bas-1}, one obtains a {\em Weyl basis} for
$A_{N-1} $:
\begin{equation}
\left\{ H_\alpha \, , \; \alpha \in \Phi \right\} \cup \left\{
  E_\alpha \, , \; \alpha \in \Delta \right\}
\label{eq:weyl-bas-2}
\end{equation}

Since $ \left\{ {\rm ad}_H \; | \; H \in \mathfrak{h} \right\} $ is a
commuting family of semisimple operators, there is a basis of $A_{N-1}
$ in which these operators are simultaneously diagonalizable.
Fix $\bar H \in \mathfrak{h} $ to be one such $N \times N $ diagonal
traceless matrices, $\bar H = {\rm diag}(\lambda_1, \ldots , \lambda_N ) $
such that $ \sum_{i=1}^N \lambda_i = 0 $ where the $\lambda_i $ are
assumed to be ordered: $\lambda_1 > \ldots > \lambda_N $.
Let $ E_{ij} $ be the matrix with 1 in the $(ij) $ slot and 0
elsewhere.
Since $ {\rm ad}_{\bar H} E_{ij} = ( \lambda_i - \lambda_j ) E_{ij} $, the
$E_{ij} $ are the root vectors and the roots are the functionals $
\alpha_{ij} =  \alpha_i - \alpha_j $ such that $ \alpha_i (\bar H) =
\lambda_i $. 
The sum $\alpha_{ij} + \alpha_{kl} $ is a root if and only if $ j=k$
or $ i = l $ (if both, then it is the zero root).
In fact, from $E_{ij} E_{kl} = \delta _{jk}E_{il} $ where $ \delta_{ij} $ is the Kroneker delta, $ [ E_{ij} , \, E_{kl} ] = \delta_{jk} E_{il} - \delta_{li}
E_{kj} $, i.e.
\begin{equation}
[ E_{ij} , \, E_{kl} ] = \begin{cases}
0 & \text{ if $ j \neq k $ and $ i\neq l $ } \\
E_{il} & \text{ if $ j=k $ }  \\
-E_{kj} & \text{ if $ i=l $} \\
E_{ii} - E_{jj} & \text{ if $ j=k $ and $ i=l $}
\end{cases}
\label{eq:weyl-comm}
\end{equation}
The roots $\alpha_{ij} $ are real and such that if $ \alpha_{ij} $ is
a root so is $ - \alpha_{ij} $. 
$ \bar H $ is a regular element if $ \alpha_{ij} \neq \alpha_{kl} $,
for indexes $ i, \, j, \, k, \, l $ such that $ ( i, \, j ) \neq (k,
\, l ) $, $ i \neq j $ and $ k \neq l $.
Thus $\bar H $ is regular if and only if $ \lambda_i - \lambda_j \neq
\lambda_k - \lambda_l $.
The fundamental roots are $ \alpha_{12} , \; \alpha_{23} , \ldots ,
\alpha_{N-1, N} $ and a basis of $ A_{N-1} $ corresponding to
\eqref{eq:weyl-bas-2} is given by 
\begin{equation}
\left\{ H_i = E_{ii} - E_{i+1, i+1} \, , \; i = 1, \ldots , N-1
\right\} \cup \left\{ E _{ij } \, , \; i, \, j = 1, \ldots , N \, , \;
  i\neq j \right\}
\label{eq:weyl-bas-3}
\end{equation}
$\mathfrak{su}(N)$ is the {\em compact real form} of $A_{N-1} $ since
it corresponds to a negative definite Killing form.
The basis of $\mathfrak{su}(N) $ corresponding to \eqref{eq:weyl-bas-2} is
\begin{equation}
\left\{ i H_\alpha \, , \; \alpha \in \Phi \right\} \cup \left\{
  X_\alpha = E_\alpha - E_{- \alpha }  \, , \; \alpha \in \Delta
  ^+\right\} \cup \left\{ Y_\alpha =i ( E_\alpha + E_{- \alpha })  \,
  , \; \alpha \in \Delta^+ \right\} 
\label{eq:sun-bas-1}
\end{equation}
or, after diagonalization of the Cartan subalgebra
\begin{equation}
\left\{ i H_i \, , \; i = 1, \ldots , N-1
\right\} \cup \left\{X_{ij} =  E _{ij }- E_{ji}  \, , \; 1 \leqslant i < j
  \leqslant N \right\}  \cup \left\{ Y_{ij} = i (  E _{ij }+ E_{ji} ) \, ,
  \; 1 \leqslant i < j   \leqslant N \right\}
\label{eq:sun-bas-2}
\end{equation}
Indeed, this skew-Hermitian basis forms a real Lie algebra as all the
structure constants are real:
\begin{equation}
\begin{split}
[ X_{ij} , \, X_{kl} ] & =  \delta_{jk} X_{il}+ \delta_{il}
X_{jk} + \delta_{jl} X_{ki} + \delta_{ik} X_{lj} \\
[ Y_{ij} , \, Y_{kl} ] & =  \delta_{jk} X_{li}+ \delta_{il}
X_{kj}+\delta_{jl} X_{ki}+\delta_{ik} X_{lj} \\
[ X_{ij} , \, Y_{kl} ] & =  \delta_{jk} Y_{il} - \delta_{il} Y_{kj} +
\delta_{jl} Y_{ik} - \delta_{ik} Y_{lj} \\
[ i H_i , \, X_{jk} ] & =  \delta_{ij} Y_{ik} - \delta_{ik} Y_{ji}  -
\delta_{i+1 , j} Y_{i+1, k } + \delta_{k, i+1} Y_{j, i+1} \\
[ i H_i , \, Y_{jk} ] & =  \delta_{ij} X_{ki} + \delta_{ik} X_{ji} +
\delta_{i+1, j} X_{i+1, k} + \delta_{k, i+1} X_{i+1, j}
\end{split}
\label{eq-su-comm}
\end{equation}

The basis \eqref{eq:sun-bas-1} corresponds to the direct sum,
orthogonal with respect to the Killing form:
\begin{equation}
\mathfrak{su}(N) = \bigoplus_{\alpha \in \Phi } i \mathbb{R} H_\alpha 
 \bigoplus_{\alpha \in \Delta^+} \mathbb{R}  X_\alpha
 \bigoplus_{\alpha \in \Delta^+}\mathbb{R}Y_\alpha
\label{eq:sun-dec-1}
\end{equation}
If $ A $ is in the Cartan subalgebra of $\mathfrak{su}(N) $ then $ A =
i H $ with $ H \in \mathfrak{h} $.
Since the values of the roots at $H$, $ \alpha (H) $, are real, $
\alpha (A) $ will be imaginary and, from \eqref{eq-su-comm},
\begin{equation}
\begin{split}
{\rm ad} _A X_\alpha & = \alpha(H) Y_\alpha \\ 
{\rm ad} _A Y_\alpha & = - \alpha(H) X_\alpha
\end{split}
\label{eq:ad-H-real}
\end{equation}
thus the vector space $\mathfrak{f}_\alpha = \mathbb{R} X_\alpha +
\mathbb{R} Y_\alpha $ is invariant for $ {\rm ad}_A $.
Furthermore, the vector spaces corresponding to the fundamental
roots are enough to generate all the $\alpha$-strings
and therefore
\begin{lemma}
\label{le:gen-su-1}
 $ \left\{ \bigoplus_{\alpha \in \Phi }
  \mathfrak{f}_\alpha \right\} _{L.A.}=\mathfrak{su}(N) $
\end{lemma}
\proof 
Similarly to \eqref{eq:weyl-comm}, in the basis
\eqref{eq:sun-bas-2} we obtain (using \eqref{eq-su-comm}):
\begin{equation}
[ \mathfrak{f}_{ij} , \, \mathfrak{f}_{kl} ] = \begin{cases}
\emptyset & \text{ if $ j \neq k $ and $ i\neq l $ } \\
\mathfrak{f}_{il} & \text{ if $ j=k $ }  \\
\mathfrak{f}_{kj} & \text{ if $ i=l $} \\
\in \mathfrak{h} & \text{ if $ j=k $ and $ i=l $}
\end{cases}
\label{eq:su-comm-2}
\end{equation}
Thus from $ \left\{  \bigoplus_{\alpha \in \Phi }
  \mathfrak{f}_\alpha \right\} $ it is possible to generate $ \left\{
  \bigoplus_{\alpha \in \Delta^+ } \mathfrak{f}_\alpha \right\} $.
Moreover, $ [ X_{i, i+1} , \, Y _{i, i+1} ] = 2 i H_i $, $ i = 1,
  \ldots, N-1 $, therefore also $i \mathfrak{h} $ is generated.
\qed

On the other hand, a proper subset of fundamental roots cannot
generate $\mathfrak{su}(N) $.
\begin{lemma}
\label{le:gen-su-2} 
If $\Phi' \subsetneq \Phi $ then
 $ \left\{ \bigoplus_{\alpha \in \Phi' }
  \mathfrak{f}_\alpha \right\} _{L.A.} \subsetneq \mathfrak{su}(N) $
\end{lemma}
\proof 
Trivial since by its very definition a fundamental root cannot be
written as a sum of other positive roots, therefore if $\bar \alpha $ is a missing
fundamental root, $ \nexists$ $ \alpha , \, \beta \in \Phi $ such
that $ [ E_\alpha , \, E_\beta ] = N_{\alpha , \, \beta} E_{\bar
  \alpha }$.
Thus $ X_{\bar \alpha} $ and $ Y_{\bar \alpha }$ are not spanned by
any bracket.
\qed

\section{Quantum control system}
\label{sec:qu-co-sys}
Consider a finite level quantum system described by a state $ | \psi
\rangle$ evolving according to the time dependent Schr\"{o}dinger equation
\begin{equation}
i \hbar  {\ket{\dot \psi(t) }} = \left( \hat H_0 + u(t) \hat H_1 \right)
\ket{\psi(t) }  
\label{eq:schrod1}
\end{equation}
where the traceless Hermitian matrices $ \hat H_0 $ and $\hat H_1 $ are
respectively the internal (or free) Hamiltonian and the external
Hamiltonian, this last representing the interaction of the system with
a single control field $ u(t)$.
In the $N$-level approximation, the state $\ket{\psi} $ is expanded
with respect to a basis of $N$ orthonormal eigenstates $ \ket{\varphi_i}$: $
\ket{\psi} = \sum_{i=1}^N c_i \ket{\varphi_i} $ where the $ c_i $ are
complex coefficients that satisfy the normalization condition $
\sum_{i=1}^N |c_i |^2 = 1 $.
If we write the initial condition of \eqref{eq:schrod1} as
$\ket{\psi_0} = \sum_{i=1}^N c_{0i} \ket{\varphi_i} $, then also the
vector $ c = [ c_1 \ldots c_N ] ^T $ satisfies a differential equation
similar to \eqref{eq:schrod1}:
\begin{equation}
\begin{split}
i \hbar  { \dot c(t) } & = \left( \tilde H_0 + u(t) \tilde H_1 \right) c(t)
\\
c(0) &=  c_0  
\end{split}
\label{eq:schrod2}
\end{equation}
where now the traceless Hermitian matrix $\tilde H _0 $ is
diagonal.
The real coefficients $ {\cal E}_i $, ${\cal E}_1 \leq \ldots \leq {\cal
E}_N $, appearing along the diagonal of $\tilde H_0 $ are eigenvalues,
$ \tilde H_0\ket{\varphi_i} = {\cal E}_i \ket{\varphi_i}$, and
represent the energy levels of the system.
If $ {\cal E}_i = {\cal E}_j $ for some $ i\neq j $, then the system is
said {\em degenerate}.
If, instead, some of the levels are equispaced, $ {\cal E}_{i} - {\cal
  E}_j =  {\cal E}_k - {\cal E}_l $ for $ (i, \, j) \neq (k, \, l ) $,
$ i \neq j $, $ k \neq l $, then the system is said to have {\em degenerate
transitions} (or resonances). 
The solution of \eqref{eq:schrod2} is $ c(t) = X(t) c(0) $ with the
unitary matrix $X(t)$ representing the time evolution operator.
If we use atomic units ($\hbar = 1 $), then instead of
\eqref{eq:schrod2} we can study the right invariant bilinear control
system evolving on the Lie group $SU(N)$ and characterized by the
skew-Hermitian vector fields $ A = - i \tilde H_0 $ and $ B=- i
\tilde H_1 $: 
\begin{equation}
\begin{split}
\dot  X(t)  & = \left(A + u(t) B\right) X(t) \qquad X (t) \in SU(N)
, \quad A, \, B \in \mathfrak{su}(N)  \\
X(0) & = I
\end{split}
\label{eq:schrod3}
\end{equation}

The system \eqref{eq:schrod3} is said (globally) {\em controllable} if the
reachable set 
\begin{quote}
  \begin{tabbing}
$ {\cal R}_{\left\{ A, \, B \right\}} = $ \= $  \left\{ \bar X \in
  SU(N) \; | \;  \right. $ there exists an admissible input $ u(\cdot ) $
  such that the integral\\ 
 \>  curve of \eqref{eq:schrod3} satisfies $ X(0)
  =I , \, X(t) = \bar X $ for some $ \left. t \geqslant 0  \right\} $
     \end{tabbing} 
\end{quote}
is all of the Lie group: ${\cal R}_{\left\{ A, \, B \right\}} = SU(N)
$.
Given (any) $ A, \, B \in \mathfrak{su}(N) $, call $ \left\{ A, \, B
\right\} _{L.A.} $ the Lie algebra generated by $ A $ and $B$ with
respect to the Lie bracket.
The literature on the subject of quantum control, see Ref.
\cite{DAlessandro2,Ramakrishna1,Albertini1}, has relied essentially on the
condition of the following Theorem (originally due to \cite{Jurdjevic3}):
\begin{theorem}
\label{thm:rank-sun}
The system \eqref{eq:schrod3} is controllable if and only if $ \left\{
  A, \, B \right\} _{L.A.} =\mathfrak{su}(N) $. 
\end{theorem}
Theorem \ref{thm:rank-sun} is a consequence of the following Lemma,
which affirms that subsemigroups of compact groups are always subgroups:
\begin{lemma}
\label{lemma-semigroup} {\rm (Lemma 1, Ch.6 of \cite{Jurdjevic1})} 
For the compact semisimple Lie group $SU(N) $
\[
{\rm cl} \left( {\rm exp} (t A , \; t< 0 ) \right) \subset {\rm cl}
\left( {\rm exp} (t A , \; t\geqslant 0 ) \right) \qquad \quad \forall
\; A \in \mathfrak{su}(N) 
\]
where ${\rm exp } \, : \, \mathfrak{su}(N) \to SU(N) $ is the Lie
group exponential map (and $ {\rm cl} $ means closure).
\end{lemma}
Consequently, the drift vector field $ A$ of \eqref{eq:schrod3} is not
hampering controllability on the large and thus Theorem
\ref{thm:rank-sun} follows \footnote{Since the controllability in this
case is obtained for ``large times'' the situation would be
different if ``small time local controllability'' were of interest,
see \cite{DAlessandro3}.}.
Furthermore, the semisimple character of $\mathfrak{su}(N) $ implies
the following:
\begin{lemma}
\label{lemma-semis-dense} {\rm (Theorem 12, Ch.6 of
\cite{Jurdjevic1})} 
The set of pairs $ A, B \in \mathfrak{su}(N) $ such that $  \left\{
  A, \, B \right\} _{L.A.} =\mathfrak{su}(N) $ is open and dense in $
\mathfrak{su}(N)$.
\end{lemma}
Putting together Theorem \ref{thm:rank-sun} and Lemma
\ref{lemma-semis-dense} then we have:
\begin{corollary}
The system \eqref{eq:schrod3} is controllable for almost all pairs $
A, \, B \in \mathfrak{su}(N) $.
\end{corollary}
In spite of the generic result above, there is still some interest in
the controllability problem, especially
\begin{itemize}
\item characterize the algebraic set in which controllability may fail
\item determining alternative procedures for testing controllability,
  other than exhaustive computation of Lie brackets.
\item find sufficient conditions for controllability based on the {\em a priori} knowledge of the vector fields $A$ and $B$.
\end{itemize}
This paper is dedicated to the last two items of the list.

\subsection{Roots and graphs}
$A$ and $B$ are expressed in terms of the components of the $
\mathfrak{su}(N) $ basis \eqref{eq:sun-bas-1} as:
\begin{eqnarray}
A & = & 
=  \sum _{\alpha \in \Phi } \alpha (\tilde H_0 )
 \left( i H _{\alpha} \right) 
\label{eq:A-decomp} \\
 B & = & 
B_0 + \sum _{\alpha \in \Gamma^+ \subseteq \Delta^+ } \left(  b_\alpha^\Re
  X_\alpha +  b_\alpha^\Im  Y_\alpha \right)  
\label{eq:B-decomp}
\end{eqnarray}
where $ B_0 \in i \mathfrak{h} $, $ b_\alpha^\Re $ and $
b_\alpha^\Im $ are real and $ \Gamma^+ \subseteq \Delta^+ $ is the subset of roots
``touched'' by $B$.

In this case, it is possible to use the connectivity properties of
the graph of $B$ to draw conclusions about controllability in the same
spirit as it is done in \cite{Gauthier1, SilvaLeite2}
for {\em normal} (or {\em split}) {\em real forms} and, more recently,
in \cite{Turinici1} for the same quantum control problem.
Consider the {\em graph} ${\cal G}_B $ associated to a square matrix $ B = [
b_{ij} ] $, i.e. the pair ${\cal G}_B = ( {\cal N}_B, \, {\cal C}_B )$
where $ {\cal N}_B $ represents a set of $N$ ordered nodes $ {\cal
  N}_B = \left\{ 1, \ldots , n \right\} $ and $ {\cal C}_B $ the set
of oriented arcs joining the nodes: $ {\cal C}_B = \left\{ (i, \, j )
  \; | \; b_{ij} \neq 0 \right\}$.
The graph $ {\cal G}_B $ is said {\em strongly connected} if for all
pairs of nodes in $ {\cal N}_B $ there exists an oriented path in
${\cal C}_B $ connecting them.
${\cal G}_B $ is strongly connected if and only if $B$ is
permutation-irreducible (P-irreducible) \footnote{A permutation matrix
  $P$ has elements that are $0$ or $1$, see \cite{Varga1}},
i.e. there exists no permutation matrix $P$ such that 
\[
P^{-1} B P = \begin{bmatrix} B_1 & \ast \\ 0 & B_2 \end{bmatrix}
\]
A square matrix is $P$-irreducible if and only if its graph does not
contain any strongly disconnected subgraph.
As long as we consider matrices $B$ that are Hermitian or
skew-Hermitian, the adjective ``strong'' (referring to the path being
oriented) is irrelevant since $ b_{ij} \neq 0 $ if and only if $
b_{ji} \neq 0 $.
\begin{remark} For $B$ Hermitian or skew-Hermitian, $ {\cal G}_B $
  connected $ \Longleftrightarrow $ ${\cal G}_B $ strongly connected.
\end{remark}
This is not anymore true if $B$ belongs to a noncompact real form.
Working with the complexification $A_{N-1}$ and considering the graphs
associated with the elements $E_\alpha $, $ \alpha \in \Delta^+ $, of
the Weyl basis \eqref{eq:weyl-bas-3}, a unique $
{\cal G}_{E_\alpha} $ is associated to each positive root.
$ {\cal G}_{E_\alpha } $ are called {\em elementary root graphs}.
If $ b_\alpha = b_\alpha^\Re + i  b_\alpha^\Im $, rewriting $B$ as 
\begin{equation}
B  =  B_0 + B_1 = B_0 + \sum _{\alpha \in \Gamma^+} \left( b_\alpha
  E_\alpha -  b_\alpha^\ast E_{-\alpha} \right)  
\label{eq:B-decomp-bis}
\end{equation}
where $ \; ^\ast $ is complex
conjugate, then the (positive) {\em root 
graph} of $B$ is ${\cal G}^+_B = \bigcup _{\alpha\in \Gamma^+} {\cal
G}_{E_\alpha } $ and $ {\cal G}_{B-B_0} $ is ``twice'' $ {\cal
G}_B^+$.

For the quantum system on $\mathfrak{su}(N) $, the roots admit the
interpretation of transitions between energy levels of the system.
According to our definitions, $ \alpha_i( \tilde H_0) = - {\cal E}_i$
and the roots are $ \alpha_{ij} = {\cal E}_j - {\cal E}_i $ ($
i< j \Rightarrow \alpha_{ij}\geqslant  0 $).
In the basis \eqref{eq:sun-bas-2}, $ b_\alpha  =
b_{ij} $, $ 1 \leqslant i < j \leqslant N $ and $B_0 $ is
simply the diagonal 
\begin{equation}
B_0 = \sum_{k=1}^{N} b_{kk} E_{kk} = \sum_{k=1}^{N-1} \left( b_{i,i} -
  b_{i+1, i+1} \right) ( i H_i )  
\label{eq:B0-1}
\end{equation}
The $b_{ii}$ correspond to loops  
on ${\cal G}_B $, i.e. to arcs beginning and ending on the same node.
Thus they are irrelevant for the connectivity property.
Equations \eqref{eq:A-decomp}-\eqref{eq:B-decomp} are then
\begin{eqnarray}
A & = & \sum_{i=1}^{N-1} i \left( {\cal E}_{i+1} - {\cal E}_i \right) H_i
\label{eq:A-decomp-ij} \\
 B & = &  B_0 + \sum_{( i, j) \in {\cal C}_B^+}  \left(  b_{ij}^\Re
  X_{ij} +  b_{ij}^\Im  Y_{ij} \right)  \label{eq:B-decomp-ij} 
\end{eqnarray}

The following lemma is the adaptation to our situation of Theorem 2
and Corollary 2 of \cite{SilvaLeite2}. 
\begin{lemma}
\label{le:gen-su-3}
$B$ is $P$-irreducible $\Longleftrightarrow $ $ \left\{
  \mathfrak{f}_\alpha , \; \alpha \in \Gamma^+ \right\}_{L.A.} =
  \mathfrak{su}(N) $
\end{lemma}
\proof
If $B$ is P-irreducible, then $ B - B_0 $ is P-irreducible and $ {\cal
  G}^+ _B $ is connected.
Therefore, every
pair of nodes $(i, \, j ) $, $ i\neq j $, can be joined by a path made
up of elementary root graphs belonging to $ {\cal G}_B^+ $, or, in
terms of roots, each positive root of $A_{N-1}$, $\alpha_{ij} \in
\Delta^+ $, can be written as a sum of the
roots of $\Gamma^+ $: $ \alpha_{ij} = \sum_{(k, l)\in {\cal C}^+_B }
\lambda_{k,l} \alpha_{k,l} $ for positive rationals $
\lambda_{kl}$.
The situation is specular for negative roots.
From the commutation relations \eqref{eq:su-comm-2}, for some vector
spaces $ \mathfrak{f}_{k_i, l_i} $ of indexes $ ( k_i, \, l_i ) \in
{\cal C}^+_B $ we have
\begin{equation}
[ \mathfrak{f}_{k_1, l_1} , \, [  \mathfrak{f}_{k_2, l_2} , \ldots [
\mathfrak{f}_{k_{m-1}, l_{m-1}} , \, \mathfrak{f}_{k_m, l_m} ] \ldots
] = \mathfrak{f}_{i, j}
\label{eq:su-comm-comm}
\end{equation}
Then also the Cartan subalgebra can be generated,
see \eqref{eq-su-comm}, and the result follows.
On the other direction, $\left\{
\mathfrak{f}_\alpha , \; \alpha \in \Gamma^+ \right\}_{L.A.} =
\mathfrak{su}(N) $ means that repeated brackets like
\eqref{eq:su-comm-comm} span all the subspaces $\mathfrak{f}_{ij} $, $
1 \leqslant i < j \leqslant N$, and therefore touch all the roots $
\alpha_{ij} \in \Delta$. 
But this is equivalent to $ {\cal G}_B $ connected.  
\qed

Lemma \ref{le:gen-su-1} and Lemma \ref{le:gen-su-2} tell us that the
condition of Lemma \ref{le:gen-su-3} is ``minimally'' satisfied by a
set of fundamental roots, although due to the nonuniqueness of the
selection of the fundamental roots, not all the $ \alpha \in \Phi $
have to be in $\Gamma ^+ $ for ${\cal G}_B$ to be connected.
\begin{corollary}
\label{cor:fund-suff1}
If $ \Phi \subseteq \Gamma^+ $ then $ B $ is $P$-irreducible.
\end{corollary}

\section{Sufficient conditions for controllability in the generic case}
\label{sec:suff-co-gen}
Considerations similar to those used in the controllability analysis
of normal real forms of classical Lie 
algebras (see \cite{Jurdjevic4,Gauthier1,ElAssoudi1,SilvaLeite2}) can be employed for our compact real form as well. 
In the case of free Hamiltonian of diagonal type, the connectivity property of
the graph of the forced term $B$ can replace the Lie algebraic rank
condition, see \cite{Gauthier1}.

\begin{lemma}
If $A$ is diagonal, a necessary condition for controllability is that
$ {\cal G}_B $ connected.
\end{lemma}
\proof
If $B$ is $P$-reducible, then there exist nontrivial invariant
subspaces of $\mathfrak{su}(N) $ that are simultaneously $A$-invariant
and $B$-invariant. Thus the system cannot be controllable.
\qed

In the case of ${\cal G}_B $ disconnected, the quantum system is
decomposable into noninteracting subsystems \footnote{In Turinici \cite{Turinici1} it is required that $
  \tilde H_1 $ is off-diagonal. 
The interpretation in terms of root decomposition offered here shows
that this assumption is irrelevant for controllability: the diagonal
terms of $ \tilde H_1 $ belong to the Cartan 
subalgebra and as such they commute with $A$.}.
  
The equivalence between $ \left\{ A , \, B \right\}_{L.A.} =
\mathfrak{su}(N) $ and ${\cal G}_B $ connected is not exact: while $
{\cal G}_B $ connected is a necessary condition for controllability,
alone it is not a sufficient condition, but requires extra assumptions
to be made on the diagonal matrix $A$.
The simplest case corresponds to the drift term $ A
$ being regular \footnote{Notice that in the case of purely imaginary eigenvalues, the concept
of {\em strongly regular} element used in \cite{Jurdjevic4} coincides
with the notion of regular element given here.} and corresponds
to all nondegenerate transitions.

\begin{theorem}
\label{thm:suff-cond-1}
Given $A$ and $B$ as in \eqref{eq:A-decomp} and \eqref{eq:B-decomp},
assume that $ {\cal G}_B $ is connected.
If $A$ is regular, then the system \eqref{eq:schrod3} is controllable.
\end{theorem}
\proof 
Since $ {\rm ad}_A $ is invariant
on each $ \mathfrak{f}_\alpha $, $ \alpha \in \Delta^+$,
 the level one bracket $ C = {\rm ad}_A \sum_{\alpha
\in \Gamma^+}\left(
  b^\Re_\alpha X_\alpha + b^\Im_\alpha Y_\alpha \right) $ allows to
reach $ \bigoplus_{\alpha \in \Gamma ^+ } 
\mathfrak{f}_\alpha $ as $ b_\alpha = b^\Re _\alpha + i
b^\Im_\alpha \neq 0 $ for all $\alpha\in \Gamma^+$.
Using an argument similar to the proof of Theorem 3 in \cite{SilvaLeite2},
we compute as many ``$A$-brackets'' (like $ [A, \, B ] $, $ [A ,\, [A \, B
] \, ] $, etc.) as the number of roots in $ \Gamma^+ $, say $2m $ (with
$m \geqslant N-1$).
For simplicity of bookkeeping, it is convenient to number roots $
\alpha $ and coefficients $ b_\alpha $ cardinally from $1$ to $ 2m$:
$ \left\{ \alpha_1 , \ldots , \alpha_{2m} \right\} = \Gamma^+ $, 
$b_\alpha = b_i $, $i=1, \ldots, 2m $ and $E_\alpha = E_i $, $i=1,
\ldots, 2m $
\[
\begin{bmatrix}
{\rm ad}_{ A} B \\
{\rm ad}_{ A}^2 B \\
\vdots \\
{\rm ad}_{ A}^{2m} B
\end{bmatrix} 
= \begin{bmatrix}
\alpha_1 b_1 & \alpha_2 b_2 & \ldots & \alpha_m b_m & \alpha_1
b_1^\ast & \ldots & \alpha_m b_m^\ast \\
\alpha_1^2 b_1 & \alpha_2^2 b_2 & \ldots & \alpha_m^2 b_m & - \alpha_1^2
b_1^\ast & \ldots & -\alpha_m^2 b_m^\ast \\
\alpha_1^3 b_1 & \alpha_2^3 b_2 & \ldots & \alpha_m^3 b_m & \alpha_1^3
b_1^\ast & \ldots & \alpha_m^3 b_m^\ast \\
 & & & & & & \\
& & & \vdots & & & \\
& & & \vdots & & & \\
& & & & & & \\
\alpha_1^{2m} b_1 & \alpha_2^{2m} b_2 & \ldots & \alpha_m^{2m} b_m & -
\alpha_1^{2m} b_1^\ast & \ldots & -\alpha_m^{2m} b_m^\ast
\end{bmatrix}
 \begin{bmatrix}
E_1 \\
E_2 \\
\vdots \\
E_m \\
E_{-1} \\
\vdots \\
E_{-m}
\end{bmatrix} = M \begin{bmatrix}
E_1 \\
E_2 \\
\vdots \\
E_m \\
E_{-1} \\
\vdots \\
E_{-m}
\end{bmatrix}
\]
$M$ can be written as
\[
M = S_1 \; 
\begin{bmatrix}
\alpha_1  & \ldots & \alpha_m & \vline & \alpha_1 & \ldots & \alpha_m \\
\alpha_1^3 & \ldots & \alpha_m^3 & \vline & \alpha_1^3
 & \ldots & \alpha_m^3 \\
& & & \vline &  & & & \\
\alpha_1^{2m-1}  &  \ldots & \alpha_m^{2m-1} & \vline &
\alpha_1^{2m-1} & \ldots & \alpha_m^{2m-1}  \\
\hline
\alpha_1^{2} &\ldots & \alpha_m^{2} & \vline & 0  & \ldots & 0
\\
\alpha_1^{4} &\ldots & \alpha_m^{4} & \vline & 0  & \ldots & 0
\\
& & & \vline&   & & & \\
\alpha_1^{2m} & \ldots & \alpha_m^{2m} & \vline &0  & \ldots & 0
\end{bmatrix} \;  S_2 \;
\begin{bmatrix}
b_1& & & \vline & & & \\
& \ddots & & \vline & & & \\
& & b_m & \vline & & & \\
\hline
& & & \vline & b_1^\ast & & \\
& & & \vline & & \ddots & \\
& & & \vline & & & b_m^\ast \\
\end{bmatrix}
\]
with 
\[
S_1 = \begin{bmatrix}
1 &       &   & \vline & -1 &        &    \\ 
  &\ddots &   & \vline &    & \ddots &    \\
  &       & 1 & \vline &    &        & -1 \\
\hline 
 & & & \vline & 2  & & \\
 & & & \vline & & \ddots & \\
 & & & \vline & & & 2 
\end{bmatrix} 
 \qquad 
S_2 =\begin{bmatrix}
1 & 0 &   &   & & \vline & 0 &   &   &    \\ 
0 & 0 &   &   & & \vline & 1 & 0 &   &   \\
0 & 1 & 0 &   & & \vline & 0 & 0 &   &  \\
0 & 0 & 0 &   & & \vline & 0 & 1 &   & \\
0 & 0 & 1 &   & & \vline & 0 & 0 &   & \\
  & & & \ddots& & \vline & & & \ddots & \\
  & & & & 1  & \vline & & & & 0 \\
  & & & & 0  & \vline & & & & 1 
\end{bmatrix} 
\]
Straightforward computations give the determinant of $M$:
\[
{\rm det} M = (-1)^{m+1} 2^m \prod_{i=1}^m \alpha_i ^m
\prod_{1\leqslant i < j \leqslant m} \left( \alpha_j^2 - \alpha_i ^2
\right)^2 \prod_{i=1}^m | b_i|^2 \neq 0 
\]
Therefore $ \left\{ A, \, B\right\}_{L.A.} \supseteq  \left\{
  \mathfrak{f}_\alpha , \, \alpha \in \Gamma^+ \right\}_{L.A.} $, and 
controllability follows from $P$-irreducibility of $B$ (Lemma
\ref{le:gen-su-3}).
\qed

A weaker property than regularity is $ B$-regularity, introduced in
\cite{SilvaLeite2}. 
\begin{definition}
Given $B$ as in \eqref{eq:B-decomp}, $A$ is said {\em $B$-regular} if
the elements $ \alpha (\tilde H _0 )$, $ \alpha \in \Gamma^+ $, are
nonzero and distinct. 
\end{definition}
Unlike regularity, which requires all roots of $ \Delta $ to be
nonnull and distinct when computed in $A$, $B$-regularity requires
the root decomposition determined by $A$ to be regular only along the
roots $ \Gamma ^+$ entering into the decomposition of $B$: $
\alpha_{ij} (\tilde H_0 )= {\cal E}_j - {\cal E}_i \neq 0 $ if $
b_{ij} \neq 0 $.  
Obviously, $A $ regular means $A$ is $B$-regular for all $ B$.

Theorem \ref{thm:suff-cond-1} is a particular case of the following:
\begin{theorem}
\label{thm:suff-cond-2}
Given $A$ and $B$ as in \eqref{eq:A-decomp} and \eqref{eq:B-decomp},
assume that $ {\cal G}_B $ is connected.
If $A$ is $B$-regular, then the system \eqref{eq:schrod3} is controllable.
\end{theorem}
\proof 
The only difference with Theorem \ref{thm:suff-cond-1} is that ${\rm
  ad}_A $ is invariant only on $ \mathfrak{f}_\alpha $, $ \alpha \in
\Gamma^+ $ (rather than $ \Delta^+ $).
However, Lemma \ref{le:gen-su-3} is true regardless of this assumption
(as it is concerned with $B$ alone) and, using the same construction of the proof of Theorem \ref{thm:suff-cond-1}, $ {\rm det}M \neq 0 $ still holds true.
\qed

An alternative extension of Theorem \ref{thm:suff-cond-1} is mentioned
in \cite{Turinici1}.
If $\Pi ^+_A $ is the set of positive roots $ \alpha ( \tilde H_0 )$ that
are regular for $A$, call $ \Theta^+_A  = \Gamma^+ \cap \Pi^+_A $ the
subset of positive regular roots of $ \Gamma^+ $ when computed in $A$
and $ \Omega^+_A $ the corresponding set of 
positive degenerate roots of $\Gamma^+$: $ \Omega^+_A = \Gamma^+
\smallsetminus \Theta^+_A$.
So $B$ splits into $ B = B_r + B_s $ with $ B_r = B_0 + \sum _{\alpha
  \in \Theta^+_A } \left(  b_\alpha^\Re X_\alpha +
  b_\alpha^\Im  Y_\alpha \right) $, the intersection of $B$ with the
regular roots, and $ B_s = \sum _{\alpha
  \in \Omega^+_A } \left(  b_\alpha^\Re X_\alpha +
  b_\alpha^\Im  Y_\alpha \right) $.

\begin{theorem}
\label{thm:suff-cond-3}
Given $A$ and $B$ as in \eqref{eq:A-decomp} and \eqref{eq:B-decomp}
assume that $ {\cal G}_B $ is connected.
If ${\cal G}_{B_r} $ is connected, then the system \eqref{eq:schrod3}
is controllable. 
\end{theorem}
\proof 
The pair $ A , \, B_r $ is such that $ A $ is $B_r$-regular. 
If $ {\cal G}_{B_r} $ is connected, Theorem \ref{thm:suff-cond-2}
applies to the pair $ (A, \, B_r )$.
If $m_r \geqslant N $ are the regular roots and $m_s $ the degenerate ones, using the argument of the proof of Theorem \ref{thm:suff-cond-1}, 
\[
\begin{bmatrix}
{\rm ad}_{ A} B \\
{\rm ad}_{ A}^2 B \\
\vdots \\
{\rm ad}_{ A}^{2m_r} B
\end{bmatrix} 
= \begin{bmatrix}
{\rm ad}_{ A} B_r \\
{\rm ad}_{ A}^2 B_r \\
\vdots \\
{\rm ad}_{ A}^{2m_r} B_r
\end{bmatrix} 
+ \begin{bmatrix}
{\rm ad}_{ A} B_s \\
{\rm ad}_{ A}^2 B_s \\
\vdots \\
{\rm ad}_{ A}^{2m_r} B_s
\end{bmatrix} 
= 
 M_r \begin{bmatrix}
E_1 \\
E_2 \\
\vdots \\
E_{m_r} \\
E_{-1} \\
\vdots \\
E_{-m_r}
\end{bmatrix}
+ M_s  \begin{bmatrix}
E_1 \\
E_2 \\
\vdots \\
E_{m_r} \\
E_{-1} \\
\vdots \\
E_{-m_r}
\end{bmatrix}
\]
where ${\rm det} M_r \neq 0 $ and $ {\rm det}M_s = 0 $ since it has linearly independent columns in correspondence of identical roots.
Thus the $B_s $ part of $B$ is not influencing the controllability property.
\qed

For controllability, it is sufficient that $ \Theta^+_A $ contains the
fundamental roots, as in this case $ {\cal G}_{B_r} $ is connected by
Corollary \ref{cor:fund-suff1}.

\begin{corollary}
\label{cor:fund-suff2}
If $ \Phi \subseteq \Theta^+_A $, then the system \eqref{eq:schrod3}
is controllable. 
\end{corollary}

Notice that the condition of Theorem \ref{thm:suff-cond-1} is the one
traditionally used in the literature to show that a generic pair of
vector fields on compact semisimple Lie algebras are generating, see
\cite{Kuranishi1,Boothby2,Bonnard1}.
For this purpose, given $A$ regular, $B$ is constructed such that ${\rm
  ad}_A $ is cyclic on $ \bigoplus_{\alpha\in \Delta^+}
\mathfrak{f}_\alpha $, for 
example by having $b_\alpha \neq 0 \; \forall \alpha\in \Delta^+
$.
This means that $ \bigoplus_{\alpha\in \Delta^+} \mathfrak{f}_\alpha $
can be spanned by ``$A$-brackets'' and thus all $\mathfrak{su}(N) $ is
generated by adding 
the elements of the Cartan subalgebra (bracketing according to the last
row of \eqref{eq:su-comm-2}).
However, here the method is not directly applicable because some of
the $b_{ij} $ elements of $B$ are allowed to be zero.
In this case, from $\bigoplus _{\alpha \in
  \Gamma^+}\mathfrak{f}_\alpha $, the missing 
subspaces must be reached by means of ``$B$-brackets'' $ [ C , \, B] $,
$ [ [ C , \, B], \, B ]  $ etc. and then their span completed by single
``$A$-brackets'' $ [A, \, [ C, \, B]  \, ]$,
$ [A, \, [ [ C , \, B], \, B ] \, ]  $, etc.

\section{Sufficient conditions for controllability in a few singular cases}
\label{sec:suff-co-sin}
The use of ``$B$-brackets'' is the {\em leit motif} of all other sufficient
conditions which are based on properties weaker than the regularity and
$B$-regularity of the diagonal vector field $A$.
From Corollary \ref{cor:fund-suff2}, these conditions correspond to at
least a pair of fundamental roots being equal.
If new diagonal terms can be provided to make up for the degenerate
transitions, then controllability can be recovered. 
From \eqref{eq:B-decomp-bis}, writing $ C $ as $ C = \sum_{\Gamma^+}
\alpha (A) \left( b_\alpha E_\alpha + b^\ast _\alpha E_{-\alpha}
\right)$, the level two bracket $ [C, \, B] $ is
\begin{equation}
\begin{split}
 D & = [C, \, B]  = [ C, \, B_0 ] + [ C , \, B_1 ] \\
& =  \left[ \sum_{\Gamma^+}\alpha (A) b_\alpha E_\alpha ,
   \, B_0 \right] +  \left[ \sum_{\Gamma^+}\alpha (A) b^\ast_\alpha
   E_{-\alpha}  ,   \, B_0 \right] \\
& +     \left[ \sum_{\Gamma^+}\alpha (A) b_\alpha E_\alpha ,
   \, \sum_{\Gamma^+} b_\alpha E_\alpha  \right] +  \left[
   \sum_{\Gamma^+}\alpha (A) b^\ast_\alpha 
   E_{-\alpha}  ,   \, \sum_{\Gamma^+} b_\alpha E_\alpha \right] \\
& -   \left[ \sum_{\Gamma^+}\alpha (A) b_\alpha E_\alpha ,
   \, \sum_{\Gamma^+} b^\ast_\alpha E_{-\alpha}  \right] 
-  \left[ \sum_{\Gamma^+}\alpha (A) b^\ast_\alpha 
   E_{-\alpha}  ,   \, \sum_{\Gamma^+} b^\ast _\alpha E_{-\alpha}
 \right]
\end{split}
\label{eq:[CB]}
\end{equation}
If $ B_0 $ is nonnull and linearly independent from $A$, it constitutes 
the simplest candidate to provide the
missing fundamental roots.
From \eqref{eq:B0-1}, the fundamental roots at $B_0 $, $ \alpha(B_0) $,
are equal to $ \beta_{i, i+1} = b_{ii} - b_{i+1, i+1} $
when expressed in the basis \eqref{eq:sun-bas-2}.
Assuming that the system is not degenerate (but
the roots can still be degenerate), equivalent versions of Theorems
\ref{thm:suff-cond-1} and \ref{thm:suff-cond-2} 
hold for $B_0$ and $C$ instead of $A$ and $B$.
\begin{theorem}
\label{thm:suff-cond-4}
If $ {\cal E}_i \neq {\cal E}_j $ for $ i \neq j $ and $ {\cal G}_B $
connected, then either of the following conditions is sufficient for
controllability of \eqref{eq:schrod3}:
\begin{enumerate}
\item $B_0 $ is regular
\item $ B_0 $ is $C$-regular
\end{enumerate}
\end{theorem}
\proof
Nondegenerate system means $ {\cal E}_i \neq {\cal E}_j $ for $ i \neq
j $, i.e. $ \alpha_{ij} \neq 0 $ $ \; \forall \; 1 \leqslant i < j \leqslant
N $.
Then in case of $ {\cal G}_B $ connected also $ {\cal G}_C $ is connected.
Therefore, $B_0 $ regular or $ C$-regular 
satisfy respectively Theorems \ref{thm:suff-cond-1} and
\ref{thm:suff-cond-2} 
for the pair $(B_0, \, C)$.
Since $ B = B_0 + B_1 $ (not $B_0$) is the available vector field, in order to complete the proof one has to verify that the $ (B_1 , \, C ) $ pair is not spoiling the maximal nonintegrability property of $( B_0 , \, C) $ i.e. $ A \oplus \left\{ B_0 , \, C \right\}_{L.A.}  = A \oplus \left\{ B , \, C \right\}_{L.A.}=\mathfrak{su}(N)  $ \footnote{Neglecting the trivial case of $A$ and $B_0$ linearly dependent.}.
But this follows from eq. \eqref{eq:[CB]}: while $ [ B_0 , \, C] \in \left\{ \mathfrak{f}_\alpha , \, \alpha \in \Gamma^+ \right\} $, whenever $ \Gamma^+  \subsetneq \Delta^+ $, $ [ B_1 , \, C] \in \left\{ \mathfrak{f}_\alpha , \, \alpha \in \tilde \Gamma^+ \subseteq \Gamma^+ \right\} $ because components along $ E_{\pm \alpha \, \pm \beta} $, $ \alpha , \, \beta \in \Gamma^+ $ are produced by the last four terms of \eqref{eq:[CB]} (see \eqref{eq:weyl-bas-1}).
If $ \tilde \Gamma^+ \subsetneq \Gamma^+ $, then $ [ B_0 , \, C] $ and $ [ B_1 , \, C] $ are automatically linearly independent.
If, instead, $\tilde \Gamma^+ = \Gamma^+ $, then from \eqref{eq:[CB]} it has to be $ \alpha + \beta \in \Gamma^+ $ for all roots $ \alpha, \, \beta \in \Gamma^+ $ such that $ \alpha + \beta $ is a root and $ \alpha b_\alpha ^\ast E_{-\alpha} b_\beta E_\beta - \beta b_\beta E_\beta b_\alpha^\ast E_{-\alpha} =0 $ for all $ \alpha, \; \beta \in \Gamma ^+ $ such that $ \alpha \neq \beta $, i.e. $ ( \alpha - \beta ) b_\alpha^\ast b_\beta [ E_{- \alpha}  , \, E_\beta ] = 0 $, that is $ \alpha = \beta $ for all $ \alpha, \; \beta \in \Gamma ^+ $ such that $ \alpha - \beta \in \Delta $.
Since $ {\cal E}_i \neq {\cal E}_j $, for $ i \neq j$, $ \alpha = \beta \neq 0 $.
Linear independence of $ [ B_0 , \, C] $ and $ [ B_1 , \, C] $ then follows also in this case from the fact that $ [ B_1 , \, C] $ has terms on the diagonal which are certainly nonnull for the case of all equal roots (they are computed in detail below, see equations \eqref{eq:dk} and \eqref{eq:dk2}), while $[ B_0 , \, C] $ is off-diagonal.
In both cases thus, the basis elements obtained from $ [ B_0 , \, C] $, $  [ B_0, \,  [ B_0 , \, C] \, ] $, etc. are not canceled by the remaining parts of the brackets $ [ B_1 , \, C] $, $  [ B_0, \,  [ B_1 , \, C] \, ] +  [ B_1, \,  [ B_0 , \, C] \, ]+  [ B_1, \,  [ B_1 , \, C] \, ] $, etc.
\qed

One can think of weakening further the hypothesis of Theorem \ref{thm:suff-cond-4} by combining together regular pieces from both $A$ and $B_0$.
To this end, analogously to what was done for the diagonal matrix $A$, call $ \Theta_B^+ $ the set of positive regular roots $ \alpha(B_0) $ of $ \Gamma^+$ and $ C_r $ the corresponding part of $C$: $ C_r = \sum _{\Theta^+_B}
\alpha (A) \left( b_\alpha E_\alpha + b^\ast _\alpha E_{-\alpha}
\right)$.
\begin{theorem}
\label{thm:suff-cond-5}
Assume $ {\cal E}_i \neq {\cal E}_j $ for $ i \neq j $ and $ {\cal G}_B $
connected.
If ${\cal G}_{B_r} \cup {\cal G}_{C_r} $ is connected then the system \eqref{eq:schrod3}
is controllable. 
\end{theorem}

\proof
In this case it is necessary to use both ``A-brackets'' for the pair $ [ A, \, B ] $ and ``B-brackets'' for $[B_0 , \, C]$.
Since ``A-brackets'' and ``B-brackets'' yield independent new generators, the proof follows by combining the arguments used in proving Theorem \ref{thm:suff-cond-3} and Theorem \ref{thm:suff-cond-4}.
\qed

As last, we treat the case of Cartan subalgebras from level two brackets of $A$ and $B$.
Since $C$ is off-diagonal, the only useful bracket in this respect is $[C, \, B]$.
By looking at the commutators for the Weyl basis \eqref{eq:weyl-bas-1},
new diagonal terms appear only on the
4th and 5th terms of the expression \eqref{eq:[CB]} for $D$.
By isolating them
\[
D_0 = - 2 \sum_{\Gamma^+} \alpha (A) | b_\alpha |^2 [ E_\alpha , \,
E_{-\alpha} ] = - 2 \sum_{\Gamma^+} \alpha (A) | b_\alpha |^2 H_\alpha 
\]
or, in terms of the basis \eqref{eq:sun-bas-2}
\[
D_0 = -2 \sum_{( i, \, j ) \in {\cal C}_B^+} \alpha _{ij} (A) | b_{ij}|^2
\left( H_i + H_{i+1} + \ldots + H_{j-1} \right) 
\] 
Thus $ D= [C, \, B] = D_0 + D_1 $ with $ D_0$ diagonal and $D_1$
off-diagonal.
It is convenient to sum over $ \Delta^+ $ rather than $ \Gamma^+$ (if
$ ( i, \, j ) \notin {\cal C}_B^+ $ then $ b_{ij} = 0 $):
\begin{eqnarray}
D_0 & = & -2 \sum_{i=1}^{N-1} \sum_{j =i+1} ^N \alpha _{ij} (A) |
b_{ij}|^2 \left( E_{ii} - E_{jj} \right) \nonumber \\
& = & 2 \sum_{k=1}^N \left( \sum_{i=1}^{k-1} \alpha_{ik} |b_{ik}|^2 -  
 \sum_{i=k+1}^{N} \alpha_{ki} |b_{ki}|^2 \right) E_{kk} 
=  \sum_{k=1}^N d_k E_{kk} 
\label{eq:dk}
\end{eqnarray}
where it is intended that $  \sum_{i=1}^{k-1} \alpha_{ik} |b_{ik}|^2 =
0 $ if $ k=1 $ and $  \sum_{i=k+1}^{N} \alpha_{ki} |b_{ki}|^2 =0 $ if
$ k =N$.
The diagonal elements $ d_k $ of $D_0$ can be expressed in terms
of the energy levels $ {\cal E}_k $ of the quantum system
\eqref{eq:schrod1} as
\begin{eqnarray}
d_k &= & 2 {\cal E}_k \left( | b_{1,k} |^2 + \ldots +  | b_{k-1,k}
  |^2 +  | b_{k,k+1} |^2 + \ldots +  | b_{k,N} |^2 \right) - \nonumber \\
& - & 2  \left(
  {\cal E}_1 | b_{1,k} |^2 + \ldots +  {\cal E}_{k-1} | b_{k-1,k}
  |^2 +   {\cal E}_{k+1} | b_{k,k+1} |^2 + \ldots +  {\cal E}_N |
  b_{k,N} |^2 \right) \label{eq:dk2}
\end{eqnarray}
from which it is straightforward to check that $ \sum_{k=1}^N d_k  =0 $ (thus that $D_0 \in \mathfrak{su}(N) $).

Now we can reformulated Theorem \ref{thm:suff-cond-4} with $D_0$ replacing $ B_0$.
\begin{theorem}
\label{thm:suff-cond-6}
If $ {\cal E}_i \neq {\cal E}_j $ for $ i \neq j $ and $ {\cal G}_B $
connected, then any of the following conditions is sufficient for
controllability of \eqref{eq:schrod3}:
\begin{enumerate}
\item $D_0 $ is regular
\item $ D_0 $ is $B$-regular
\item $ D_0 $ is $C$-regular
\end{enumerate}
\end{theorem}
\proof
The proof is completely analogous to that of Theorem \ref{thm:suff-cond-4}, with the extra simplification that now $ D_1 $, when $ \neq 0 $, is linearly independent from both $B $ and $C$ regardless of the regularity of the roots computed in $A$ and $B$ respectively.
\qed

The practical situations in which Theorems \ref{thm:suff-cond-4}-\ref{thm:suff-cond-6} apply are when the system has resonant modes (which correspond to degenerate roots).
The extreme case is when $ {\cal E}_{i+1} -{\cal E}_{i} = {\rm const} $ $ \forall \; i =1 , \ldots n-1 $ (all equally spaced energy levels).

Often a case-by-case analysis provides less strict sufficient conditions than those discussed in this paper.
As an example, consider the completely harmonic system mentioned above. If $ b_{i, \, j } = 0 $ for $ j \neq i \pm 1 $ (dipole approximation) \footnote{Since  $ \Gamma ^+ = \Phi ^+ $ , such pair of vector fields $A$ and $B$ is {\em minimal} among the generating pairs, in the sense of Lemma \ref{le:gen-su-2}, which implies that methods like Theorem \ref{thm:suff-cond-3} are inapplicable.} and $ b_{i, i\pm 1} = b_{i, i\pm 1}^\Im$, we are exactly in the situation described in \cite{Schirmer2}, Section 2.3.
In this case, all the fundamental roots are equal, $ \alpha_{i, \, i+1} = \mu $, $ i=1, \ldots N-1 $, $ B = \sum_{i=1}^{N-1} b_{i, i+1} Y_{ i, i+1 } $ ($B_0 =0$) and $ C = \mu \sum_{i=1}^{N-1} b_{i, i+1} X_{ i, i+1 } $.
Thus $ D = -2 \mu \sum_{i=1}^{N-1} b_{i, i+1}^2 (i H_{ i }) $  and the ``new fundamental roots'' are  $ -2 \mu ( b_{i, \, i+1 }^2 -  b_{i+1, \, i+2} ^2) $, which are not necessarily distinct for $ i= 1, \ldots , N-1 $. Thus Theorem \ref{thm:suff-cond-6} needs not be verified. For the same system, Theorem 3 of \cite{Schirmer2} provides an alternative sufficient condition that is weaker than any of the items of Theorem \ref{thm:suff-cond-6}, obtained by making use of the special structure of the system to compute the a full set of generating brackets explicitly. 
The drawback of this method is that whenever the structure of $B$ or the values of the fundamental roots are modified, the algorithm needs to be redesigned.

 \bibliographystyle{abbrv}

\end{document}